\pdfoutput=1
\documentclass[11pt,a4paper]{article}

\usepackage[utf8]{inputenc}
\usepackage{amsmath}
\usepackage{amsfonts}
\usepackage{amssymb}
\usepackage{graphicx}
\usepackage{subcaption}

\usepackage{jheppub}
\usepackage{soul}
\usepackage{times}
\sethlcolor{green}
\usepackage{epstopdf}

\begin{document}

\author{A. Rubin\footnote{Corresponding author.}, L. Arazi, S. Bressler, L. Moleri, M. Pitt, and A. Breskin}
\affiliation{Department of Particle Physics and Astrophysics, \\
Weizmann Institute of Science, 76100 Rehovot, Israel \\
E-mail: adam.rubin@weizmann.ac.il}
\title{First studies with the Resistive-Plate WELL gaseous multiplier}
\abstract{We present the results of first studies of the Resistive Plate WELL (RPWELL): a single-faced THGEM coupled to a copper anode via a resistive layer of high bulk resistivity. We explored various materials of different bulk resistivity ($10^9 - 10^{12} \; \Omega \text{cm}$) and thickness ($0.4 - 4$ mm). Our most successful prototype, with a 0.6 mm resistive plate of $\sim 10^9 \; \Omega$cm, achieved gains of up to $10^5$ with 8 keV x-ray in Ne/5\%CH$_4$; a minor 30\% gain drop occurred with a rate increase from $10$ to $10^4$ Hz/mm$^2$. The detector displayed a full ``discharge-free'' operation---even when exposed to high primary ionization events. We present the RPWELL detector concept and compare its performance to that of other previously explored THGEM configurations---in terms of gain, its curves, dependence on rate, and the response to high ionization.  The robust  Resistive Plate WELL concept is a step forward in the Micro-Pattern Gas-Detector family, with numerous potential applications.}
\maketitle

\section{Introduction}

Future high-energy experiments present increasingly growing demands in terms of detector performance, in particular with respect to rate capability and stability. Gas–avalanche detector technologies are continuously developed to meet these challenging requirements. Modern Micro-Pattern Gaseous detectors (MPGDs) are known for their high rate capability. However, unlike wire chambers which display gain saturation effects, they suffer from occasional discharges— mainly caused by highly ionizing background events; these may be Minimally Ionizing Particles (MIPs) in detectors set to record single-photoelectrons (e.g. RICH) or hadron-induced nuclear reaction products in detectors set for recording MIPs \cite{nanjo_neutron_2005,adam_effect_2005}. Such occasional discharges can potentially damage the readout electronics and the detector electrodes: introducing dead-time and affecting detection efficiency.

Numerous attempts have been made to limit the effect of discharges in MPGDs. Previous efforts focused on cascading several multipliers, e.g. cascaded Gas Electron Multipliers (GEMs \cite{sauli_gem:_1997,sauli_imaging_2007}), THick-GEMs (THGEMs \cite{chechik_thick_2005}), or hybrid multipliers \cite{duval_hybrid_2012} spreading the avalanche charge and preventing discharge propagation within the cascade. More recently resistive films have been employed in MPGDs, with the purpose of damping the energy of occasional discharges. Examples include: resistive MICROMEGAS \cite{alexopoulos_spark-resistant_2011, wotschack_development_2012}, resistive GEMs \cite{oliveira_first_2007} ,  and resistive MicroDot \cite{fonte_development_2012} detectors. Other methods of resistive MPGDs are described in \cite{peskov_advances_2009,peskov_advances_2012} and references therein.

Among other MPGDs, THGEM-based detectors with resistive electrodes are also being considered and investigated. The THGEM is an electron multiplier in which avalanche multiplication develops within sub-millimeter diameter holes, mechanically drilled in a standard two-sided copper-clad printed circuit board (PCB). Ionization electrons, induced by incoming radiation in a conversion/drift gap, are focused into the holes and multiplied in an avalanche process under the high electric field set by the potential difference between the THGEM faces; very large gains, exceeding $10^6$, can be reached by cascading a few THGEM elements \cite{shalem_advances_2006-1, cortesi_thgem_2009}. The reader is referred to \cite{breskin_concise_2009} for a review on THGEM principles, properties and applications. Initial works with resistive films involved replacing the THGEM’s metallic electrodes with resistive ones \cite{di_mauro_development_2007}. Lately, closed-geometry THGEM-based structures (similar in geometry to the detectors described in \cite{bartol_c..t._1996,bellazzini_well_1999,rocco_development_2010,alfonsi_performance_2009}) with resistive anodes have been proposed and investigated by our group, both in laboratory studies and in test beams. These WELL structures comprise of a single-faced THGEM, copper-clad on its top side only, mounted directly on top a resistive film deposited on a thin insulating sheet---similar to the Resistive Blind THGEM proposed in \cite{alfonsi_performance_2009}. The resistive layer, of typical surface resistivity in the range $1-20$ M$\Omega$/sq, is prepared by spraying a mixture of graphite particles and epoxy on a $0.1-0.2$ mm thick FR4 sheet, using a method described for Thin Gap Chambers (TGC) manufacture \cite{amram_position_2011}. Two variants of this structure have been investigated: the resistive WELL (RWELL) and segmented resistive WELL (SRWELL). The response of these detectors (with 1 cm$^2$ pads) to MIPs is described in \cite{arazi_beam_2013, bressler_beam_2013}. In particular, the SRWELL yielded efficiencies close to unity, with inter-pad multiplicity as low as 1.1, and a rate capability of $\sim 10^4$ Hz/mm$^2$ \cite{arazi_laboratory_2013}. However, despite the discharge damping observed, the epoxy/graphite resistive layers suffer from two primary drawbacks: the surface resistivity is limited to $\sim 20$ M$\Omega$/sq (higher resistivity leads to inhomogeneous layers) and the transverse evacuation of charges (along the resistive film) leads to significant pad-multiplicity; the latter was solved in the SRWELL by segmenting the resistive layer with thin copper gridlines \cite{arazi_thgem-based_2012,arazi_laboratory_2013}.

In this work we introduce the Resistive Plate WELL (RPWELL). It involves coupling a single-faced THGEM to a layer of high bulk resistivity and capacitively reading the signal from a metallic anode in direct contact with the resistive plate (figure \ref{fig:rpwell_configuration}). The RPWELL combines the properties of THGEMs and Resistive Plate Chambers (RPCs, \cite{santonico_development_1981}). The latter employ anodes of highly resistive bulk materials ($\sim 10^{10} - 10^{12} \; \Omega$cm), that fully damp sparks but cause rate limitations; new ceramics \cite{laso_ceramic_2012, naumann_ceramics_2011, garcia_extreme_2012} and doped glass \cite{wang_development_2010} of lower resistivity values ($\sim 10^7 - 10^{10} \; \Omega$cm), permit reaching rate capabilities of up to $10^3$ Hz/mm$^2$ (for a recent review on RPCs see \cite{gonzalez-diaz_challenges_2013} and references therein). The RPWELL has two potential advantages compared to what has been previously attempted with the RWELL and SRWELL: first, the higher resistivity should provide superior discharge damping; second, transporting the accumulated charge through the layer (as opposed to transversely) should naturally lead to a lower avalanche-induced charge-spread. Like an RPC, the RPWELL has the potential to quench sparks and thus extend the dynamic range of the detector. Bashkirov et al. \cite{bashkirov_novel_2009} demonstrated an ion counter with a similar configuration, using highly resistive glass. Similarly, coupling a THGEM to a glass anode was suggested in \cite{di_mauro_new_2006}, however no results were published to the best of our knowledge.

We performed a series of systematic studies of the RPWELL, coupling single-faced THGEM electrodes to plates of various resistive materials. Since the use of high bulk-resistivity anodes can naturally lead to reduced rate capability, one of our primary objectives was to investigate the possible tradeoff between discharge-damping and rate capability of the new structure. We focused on measuring the RPWELL gain, its dependence on the radiation rate, and the stability of the detector in the presence of high primary charge for several resistive anodes, spanning a range of $\sim 10^{10} - 10^{12} \; \Omega$cm, with varying thickness. We compared these to the response of a THGEM with an induction gap, a THick WELL (THWELL) (non-resistive WELL), and an RWELL. We present the results and suggest a thin, simple, high dynamic-range detector.

\section{Experimental setup and methodology}

\begin{figure}
\begin{center}
\includegraphics[width=1\textwidth]{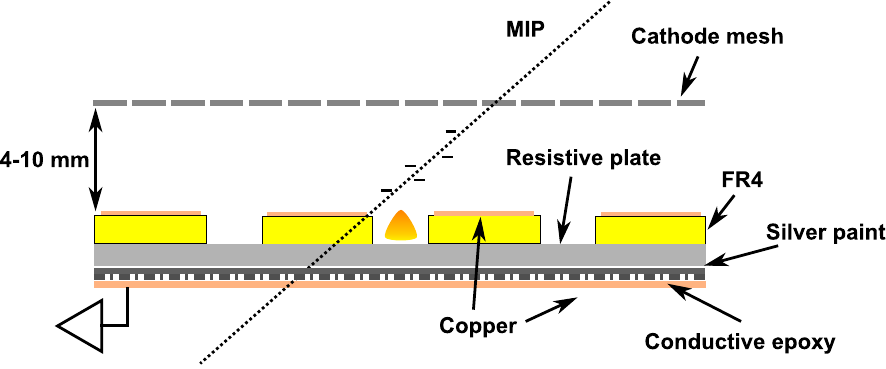}
\end{center}
\caption{The Resistive-Plate WELL (RPWELL) configuration with a resistive anode and a readout electrode. The WELL, a single-faced THGEM, is coupled to a copper anode via a resistive plate. Charges are collected from the copper anode. In some experiments the WELL was directly coupled to the metal anode.}
\label{fig:rpwell_configuration}
\end{figure}

In the experiments presented here, a single-faced THGEM was used with hole diameter d = 0.5 mm, hole pitch a = 1 mm,  thickness t = 0.8 mm, and rim etched around each hole of h = 0.1 mm. The RPWELL (figure \ref{fig:rpwell_configuration}) consists of a WELL electrode coupled to the readout anode via a high bulk-resistivity plate ($\sim 10^9-10^{12} \; \Omega \text{cm}$).

The resistive materials used in this work are listed in table \ref{tab:materials}. They were machined to about $ 30 \times 30$ mm$^2$, to match the THGEM electrode size. They were coated with a conductive paint (type: Demetron Leitsilber 200) on one side and glued with conductive epoxy (type: EPO-TEK H21D) to a copper-clad FR4 plate.  The VERTEC 400 resistive glass had aluminum evaporated onto its rear face; it was then glued with conductive epoxy to the anode plate. The $1 \; \text{M} \Omega / \text{sq}$ resistive film, used in the RWELL for comparison purposes, was prepared by spraying a epoxy/carbon film onto a 100 $\mu$m thick layer of FR4 \cite{amram_position_2011}; the electric contact was made with copper tape.

\begin{table}
\caption{The resistive materials used in this work.}
\begin{center}
\begin{tabular}{l c c}
\hline
Material & Dimensions [mm] & Resistivity \\
\hline
VERTEC 400 glass & $36 \times 31 \times 0.4$ & $8\times 10^{12} \; \Omega \text{cm}$ \\
HPL Bakelite & $29 \times 29 \times 2$ & $2\times 10^{10} \; \Omega \text{cm}$ \\
Semitron ESD 225 & $30 \times 30 \times 0.6-4$ & $2\times 10^{9} \; \Omega \text{cm}$ \\
Epoxy/carbon film & $30 \times 30 \times 0.1$ & $1 \; \text{M}\Omega / \text{sq}$ \\
\hline
\end{tabular}
\end{center}
\label{tab:materials}
\end{table}

The bulk resistivity of the materials was measured (table \ref{tab:materials}) by painting samples with a conductive paint on both sides. They were then pressed between two copper-clad FR4 electrodes and biased with a CAEN N471 power supply via a 22 M$\Omega$ resistor. The current was then monitored with a Keithley 610CR Pico ammeter as a function of the voltage. This method was used following \cite{meghna_measurement_2012},  however in the latter work the electrodes were not painted with conductive paint. Ohmic behavior was observed; the resistance was found with a linear fit, and the bulk resistivity was calculated using

\begin{equation}
R=\rho \frac{L}{A} ,
\end{equation}

\noindent where R is the resistance measured, L is the thickness of the material, A is its area, and $\rho$ is the bulk resistivity.

\begin{figure}
\begin{center}
\includegraphics[width=0.7\textwidth]{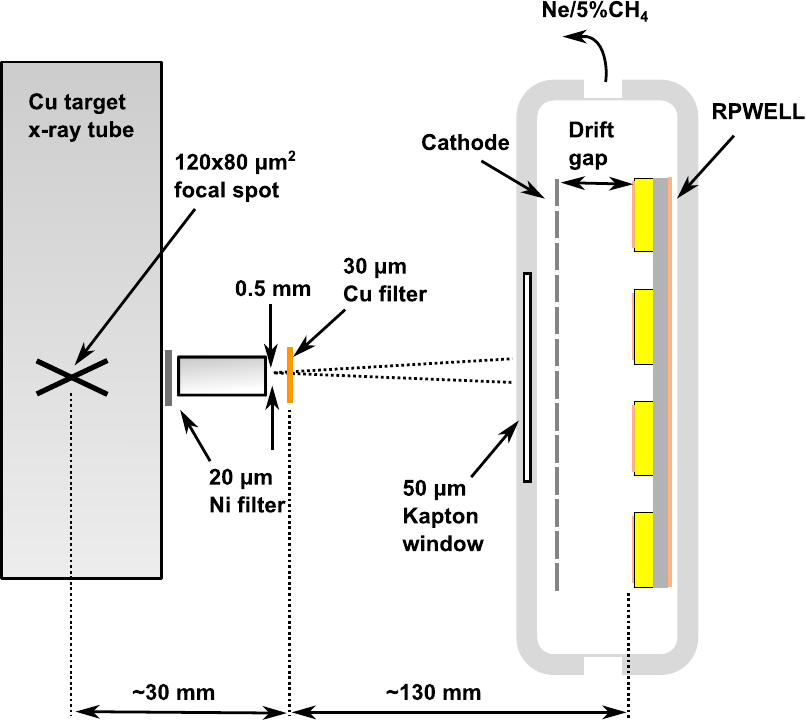}
\end{center}
\caption{Irradiation  setup. X-rays from the tube are collimated and filtered; the x-ray beam impinges on the detector through a thin Kapton window. The detector shown here is a RPWELL, preceded by an absorption drift gap.}
\label{fig:beam_configuration}
\end{figure}

The RPWELL (figure \ref{fig:rpwell_configuration}) and the other detector prototypes were assembled in an aluminum chamber, continuously flushed with 1 atm of $\text{Ne}/5\%\text{CH}_4$. The experimental setup is shown in figure \ref{fig:beam_configuration}. The detector was irradiated, through a 50$\mu$m Kapton window, with 8 keV x-rays (collimated to $0.5-5$ mm diameter, according to the experiment) through a 20 $\mu$m nickel and 30 $\mu$m copper filter (figure \ref{fig:beam_configuration}). The induced signal were recorded with an Ortec 125 charge-sensitive preamplifier (time constant $\sim 5$ ms) connected to an Ortec 570 linear amplifier set to 3 $\mu$s shaping time. The electrodes were biased with a CAEN N1471H power supply via a low-pass filter. The current and voltage from the power supply were monitored with a National Instruments DAQ analog signal digitization board NI-USB 6008. The signals were sampled at 50 Hz and recorded using National Instruments Signal Express software \cite{national_instruments_labview_2012}. The different detector prototypes were assembled with $4-10$ mm drift gaps. Gain curves and pulse shapes were recorded at rates of $\sim 10^{-2}-10^3 \; \text{Hz/mm}^2$, keeping in all measurements a drift-field value E$_{\text{drift}}=0.5$ kV/cm.

\subsection{Gain dependence on rate}

In this set of measurements, the detector (figure \ref{fig:rpwell_configuration}) was assembled  with a 10 mm drift gap. It was irradiated with a 1 mm diameter 8 keV x-ray beam. The rate was raised from  $\sim 1$ to $10^4 \;  \text{Hz/mm}^2$ by increasing the current on the x-ray tube, and by removing copper filters (figure \ref{fig:beam_configuration}). The gain was estimated by determining the centroid of the recorded spectrum; its drop with the rate, shown below in section \ref{subsec:results_rate_dependence}, depended on the detector configuration. The measured gain returned to its original value when the rate was reduced---only after several hours (as discussed in section \ref{sec:discussion}); therefore care was taken to perform all measurements with the same methodology and under the same conditions. This was the case for the RPWELL (with the resistive anode), as well as for the THWELL (with the copper anode) and RWELL (with the surface resistive film).

\subsection{Response to highly ionizing events}

\begin{figure}
\begin{center}
\includegraphics[width=0.8\textwidth]{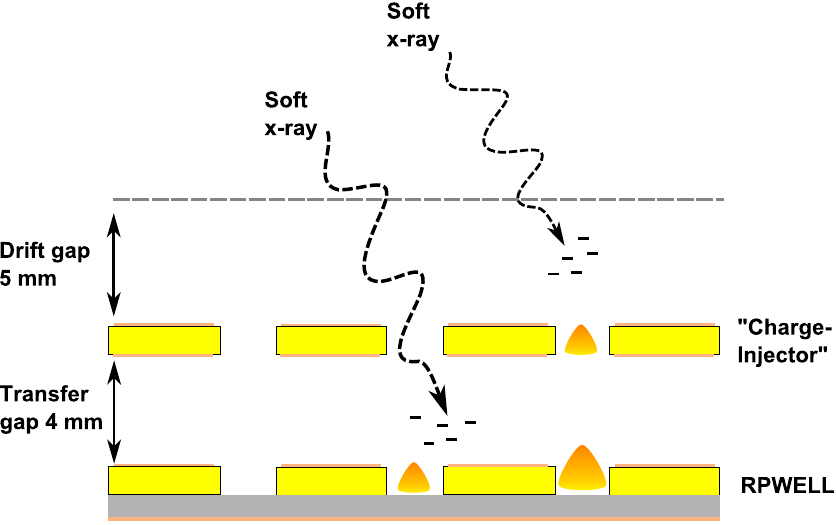}
\end{center}
\caption{Charge-injector configuration. A THGEM is used for pre-amplification of x-ray induced charges, to yield and inject a controlled number of electrons into the investigated detector (here an RPWELL), mimicking the presence of highly ionizing events in normal operation conditions. The recorded events are either those converted in the drift gap (multiplied by the two elements) or in the transfer gap (multiplied only by the second element).}
\label{fig:injector_configuration}
\end{figure}

\begin{figure}
\begin{center}
\includegraphics[width=0.5\textwidth]{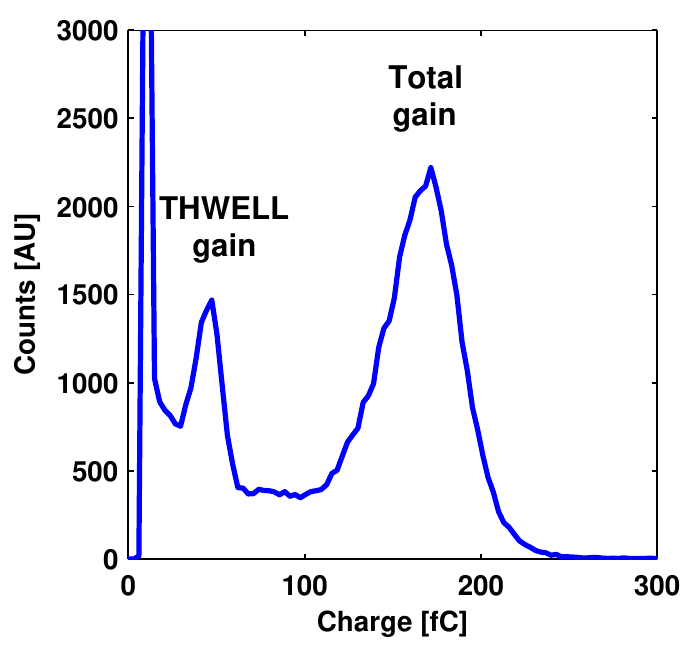}
\end{center}
\caption{Example of a spectrum recorded with 8 keV x-rays in the THWELL detector, preceded by an injector (of figure \protect\ref{fig:injector_configuration}). The two peaks correspond to events multiplied only in the THWELL (small charge) and to those that were pre-amplified by the THGEM injector (large charge). THWELL gain $4\times 10^3$, injector gain 3.6, gas Ne/5\%CH$_4$, rate $\sim 10^{-2}$ Hz/mm$^2$. }
\label{fig:injector_spectrum}
\end{figure}

In order to investigate the response of the detector to highly ionizing events we used the ``charge-injector'' method discussed in \cite{moleri_investigation_2013}. A pre-amplification THGEM (``injector'') stage was used (figure \ref{fig:injector_configuration}) to control the number of x-ray induced electrons injected into the investigated detector configuration---mimicking highly ionizing events which deposit a large amount of charge within the conversion volume. The pre-amplification THGEM had a hole diameter d = 0.5 mm, hole pitch a = 1 mm, rim h = 0.1 mm and thickness t = 0.4 mm. A 5 mm long drift gap and a 4 mm long transfer gap preceding the ``injector'' and the investigated detector respectively, had electric fields set to $\text{E}_{\text{drift}} = \text{E}_{\text{trans}} = 0.5 \; \text{kV/cm}$. Two peaks were formed in the recorded spectra (figure \ref{fig:injector_spectrum}): (1) a high-amplitude peak due to conversion in the drift gap, with the resulting electrons multiplied by the injector and then by the investigated detector (e.g. an RPWELL in figure \ref{fig:injector_configuration}), and (2) a low-amplitude peak due to conversion in the transfer gap, with the electrons multiplied solely by the RPWELL. Low charge-injector gain was calibrated by taking the ratio of the positions of the two peaks in the spectrum (figure  \ref{fig:injector_spectrum}). High injector gain---where the low peak was washed out---was estimated by using the top RPWELL electrode as an anode of the charge-injector. The investigated detector was then set to an effective gain of $ 5 \times 10^3$, irradiated with 8 keV x-rays at a conversion rate of $\sim 5 \times 10^{-1} \; \text{Hz/mm}^2$, and the injector voltage was raised in 25 V steps.

\section{Results}
\subsection{Pulse shapes and gain curves}
\label{sec:results_pulse_shapes}

\begin{figure}
\centering
\includegraphics[width=0.45\textwidth]{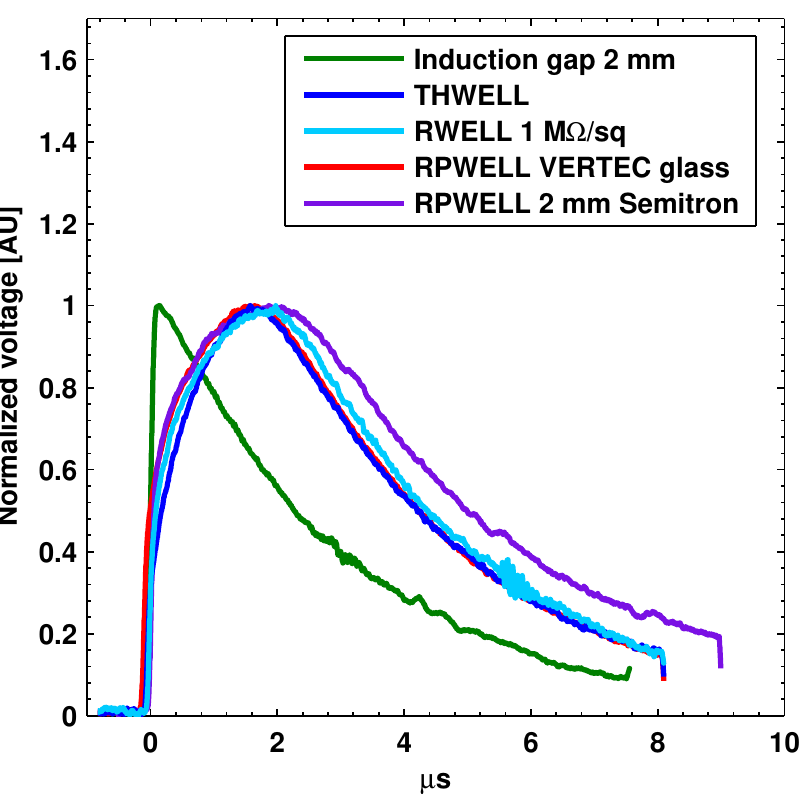}
\caption{Charge-signals recorded with an Ortec 125 charge-sensitive pre-amplifier from the anode of different detector configurations in Ne/5\%CH$_4$: THGEM with 2 mm induction gap,  THWELL, RWELL and RPWELL with VERTEC glass and Semitron resistive plates (table \ref{tab:materials}). In all cases the THGEM electrode had the same parameters, described in the text. For comparison the signals were normalized to their maxima.}
\label{fig:pulse_shapes}
\end{figure}

\begin{figure}
\centering
\begin{subfigure}{0.4\textwidth}
\includegraphics[width=\textwidth]{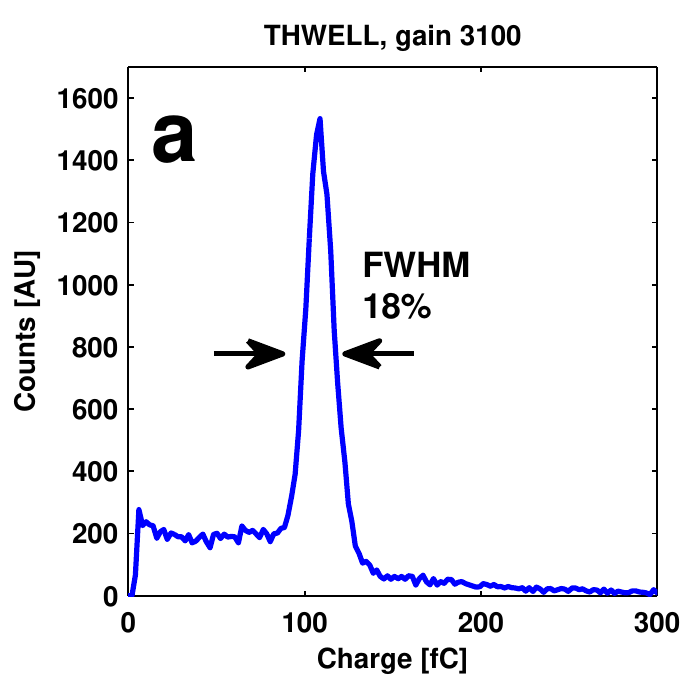}
\end{subfigure}
\begin{subfigure}{0.4\textwidth}
\includegraphics[width=\textwidth]{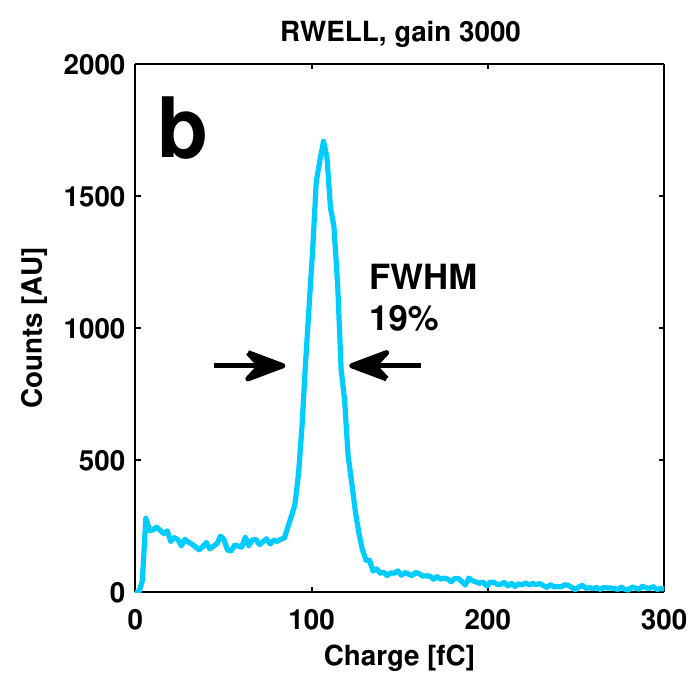}
\end{subfigure}
\begin{subfigure}{0.4\textwidth}
\includegraphics[width=\textwidth]{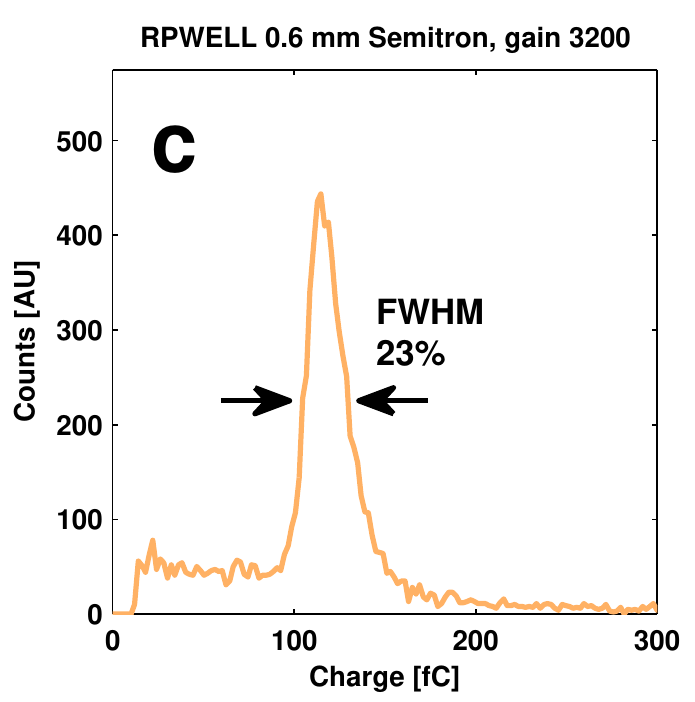}
\end{subfigure}
\begin{subfigure}{0.4\textwidth}
\includegraphics[width=\textwidth]{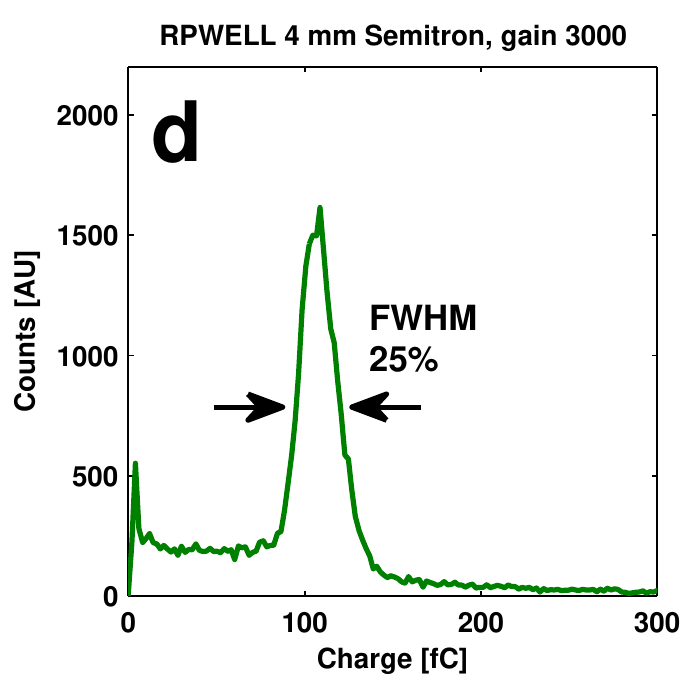}
\end{subfigure}
\caption{Pulse-height spectra acquired in different WELL detector configurations, in Ne/5\%CH$_4$, at a gain of $\sim 3 \times 10^3$ with 8 keV x-rays at $\sim 3 \times 10^2 \; \text{Hz/mm}^2$; the detectors had 10 mm drift gaps, $\text{E}_{\text{drift}}=0.5 \; \text{kV/cm}$. a) THWELL b) RWELL 1 M$\Omega/\text{sq}$ c) RPWELL with a 0.6 mm thick Semitron anode d) RPWELL with a 4 mm thick Semitron anode (table \ref{tab:materials}). The FWHM values are provided in the figures.}
\label{fig:rpwell_spectra}
\end{figure}

Figure \ref{fig:pulse_shapes} shows a comparison between anode signals recorded on a digital oscilloscope (Tektronix TDS3052)  with an Ortec 125 charge-sensitive preamplifier coupled to the detector’s anode (see figure \ref{fig:rpwell_configuration}). The figure shows signals from different detector configurations: a THWELL (coupled to a metal anode) and resistive-anode configurations, RWELL and RPWELL (the latter with two different resistive-plate materials; see table \ref{tab:materials}); a pulse from a THGEM with an induction gap is shown as well. The slower signals ($\sim 2 \; \mu$s rise time) recorded in the WELL configurations, compared to the THGEM with induction gap, are due to the avalanche-ion drift within the holes---absent with the induction gap (sensitive to avalanche electrons only). Figure \ref{fig:rpwell_spectra} shows x-ray induced pulse-height spectra acquired in different detector configurations at gains of $\sim 3 \times 10^3$ and rates of $\sim 3\times 10^2$ Hz/mm$^2$. The electronic noise of the amplification chain was similar in all of the measurements. The THWELL and RWELL (figures \ref{fig:rpwell_spectra}a and \ref{fig:rpwell_spectra}b) yielded distributions with FWHM 18\% and 19\% respectively; the RPWELL with 0.6 mm thick and 4 mm thick Semitron anodes yielded somewhat broader distributions of 23\% and 25\% (FWHM) respectively (figures \ref{fig:rpwell_spectra}c and \ref{fig:rpwell_spectra}d).

\begin{figure}
\centering
\begin{subfigure}{0.4\textwidth}
\includegraphics[width=\textwidth]{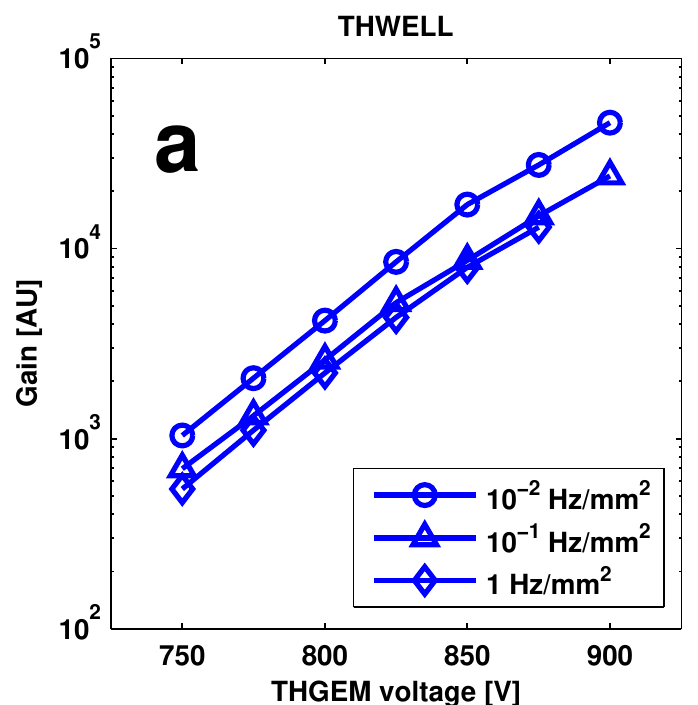}
\end{subfigure}
\begin{subfigure}{0.4\textwidth}
\includegraphics[width=\textwidth]{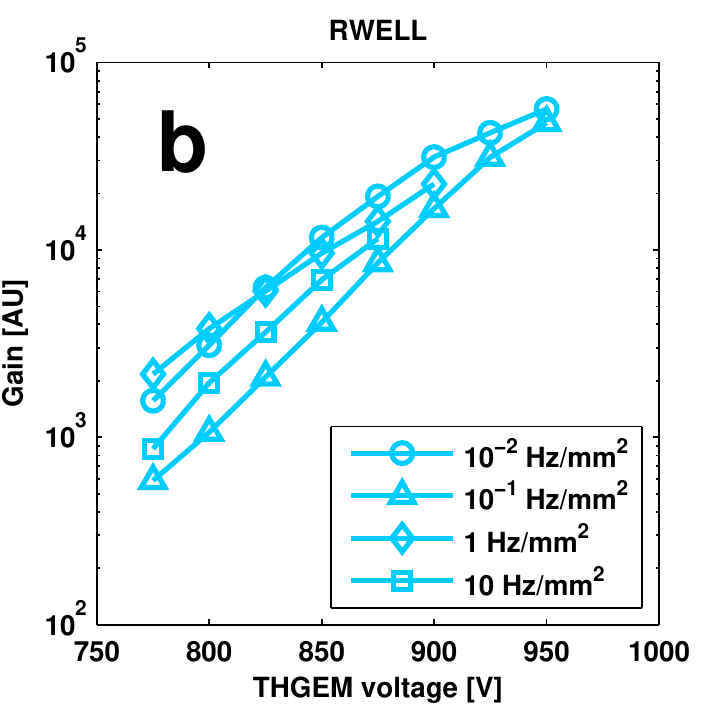}
\end{subfigure}

\begin{subfigure}{0.4\textwidth}
\includegraphics[width=\textwidth]{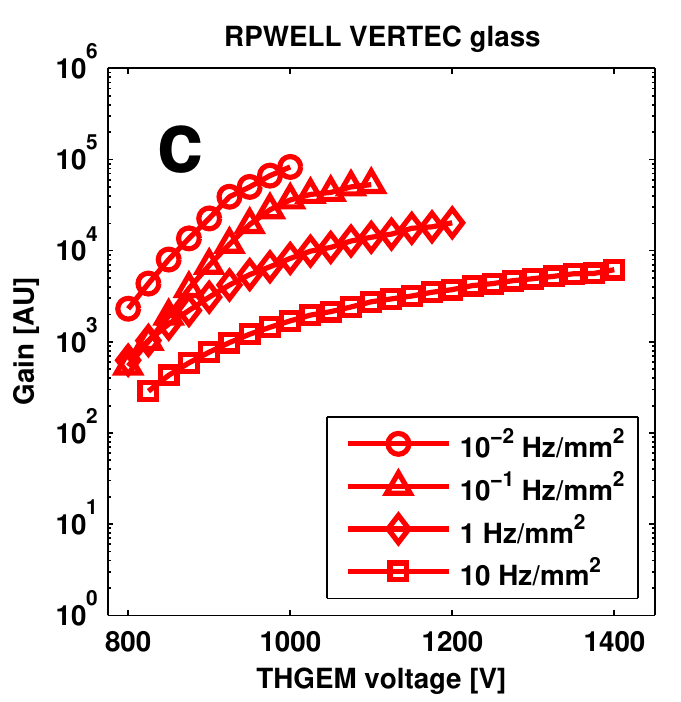}
\end{subfigure}
\begin{subfigure}{0.4\textwidth}
\includegraphics[width=\textwidth]{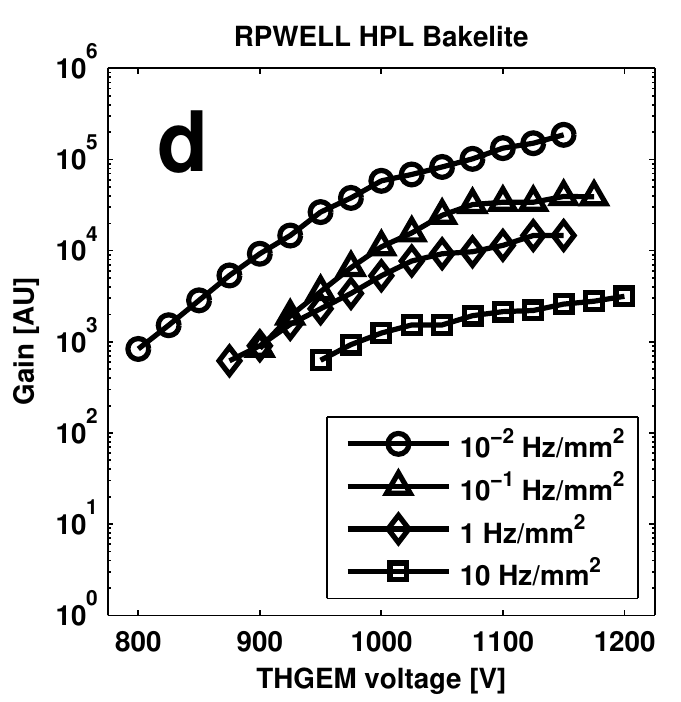}
\end{subfigure}

\caption{Gain curves recorded in Ne/5\%CH$_4$ with 8 keV x-rays at $\sim 10^{-2}-10 \; \text{Hz/mm}^2$ in different WELL-detector configurations with  $\text{E}_{\text{drift}}=0.5 \; \text{kV/cm}$ over a 4 mm drift gap. a) THWELL, b) RWELL 1 M$\Omega/\text{sq}$, c) RPWELL with VERTEC glass anode d) RPWELL with HPL Bakelite anode. The measurements with the THWELL and RWELL were terminated due to the onset of occasional discharges. The RPWELL detectors displayed an enhanced ``activity'' in the current monitor, but no discharges developed.}
\label{fig:rpwell_gain_curves}
\end{figure}

\begin{figure}
\centering
\begin{subfigure}{0.4\textwidth}
\includegraphics[width=\textwidth]{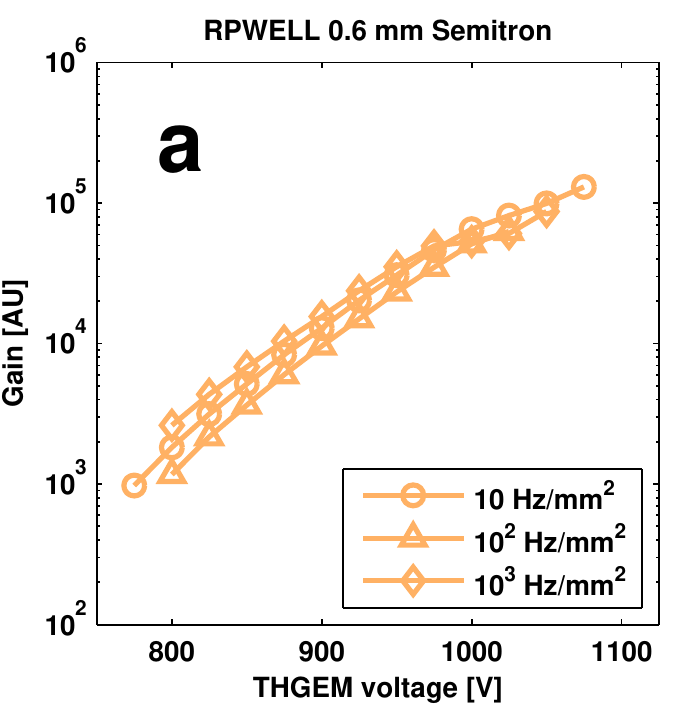}
\end{subfigure}
\begin{subfigure}{0.4\textwidth}
\includegraphics[width=\textwidth]{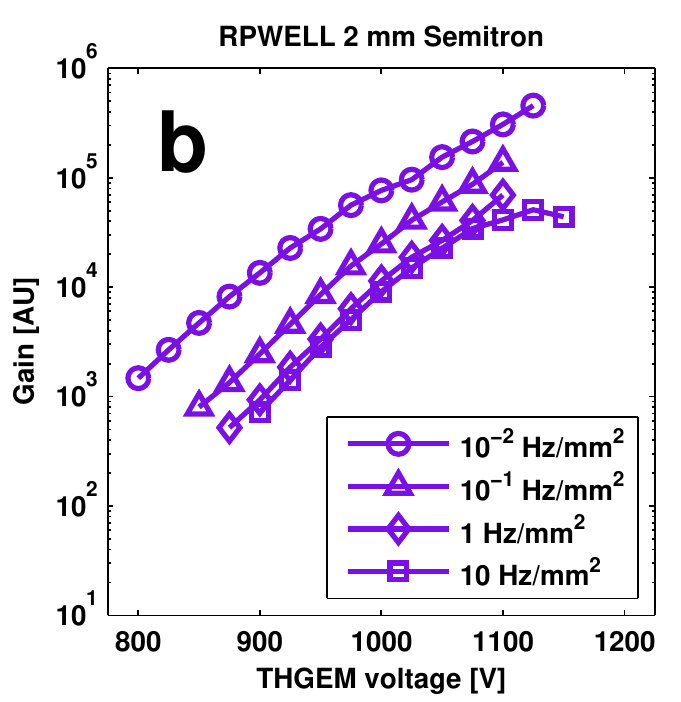}
\end{subfigure}
\begin{subfigure}{0.4\textwidth}
\includegraphics[width=\textwidth]{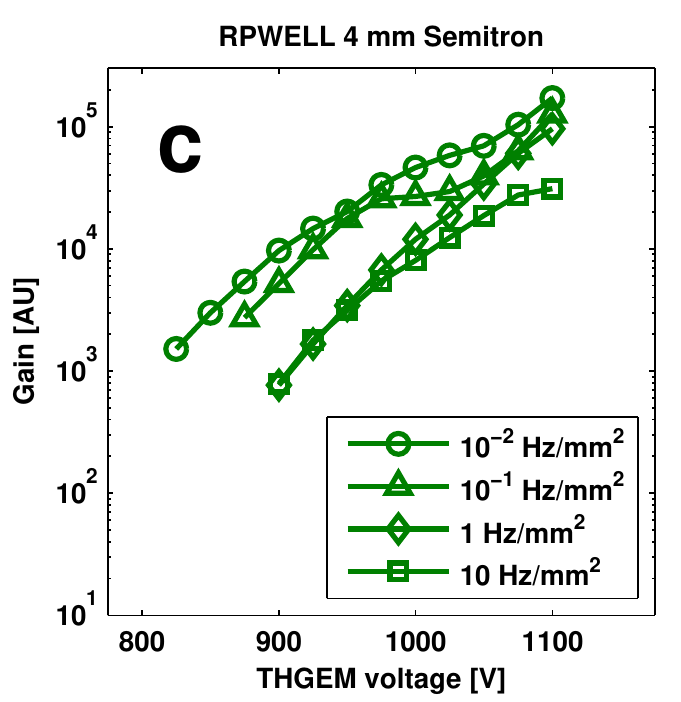}
\end{subfigure}
\caption{Gain curves recorded in Ne/5\%CH$_4$ with 8 keV x-rays at $\sim 10^{-2}-10^3 \; \text{Hz/mm}^2$ in an RPWELL-detector (figure \protect\ref{fig:rpwell_configuration}) with  Semitron ESD 225 plate anodes of different thickness and drift-gap values: a) 0.6 mm thick plate, 10 mm drift gap b) 2 mm thick plate, 4 mm drift gap c) 4 mm thick plate, 4 mm drift gap. E$_{\text{drift}} = 0.5$ kV/cm. The measurements were terminated due to an enhanced ``activity'' in the current monitor, but no discharges developed (as discussed in section \protect\ref{sec:discussion}).}
\label{fig:gain_curves_semitron}
\end{figure}

Figures \ref{fig:rpwell_gain_curves} and \ref{fig:gain_curves_semitron} show gain curves acquired with  different detector configurations (see figure \ref{fig:rpwell_configuration}) at various counting rates. While figures \ref{fig:rpwell_gain_curves}a and \ref{fig:rpwell_gain_curves}b show a rather exponential gain rise with the high voltage in the THWELL and RWELL, in the RPWELL with VERTEC glass and the HPL Bakelite (figures \ref{fig:rpwell_gain_curves}c and \ref{fig:rpwell_gain_curves}d, respectively)  the gain tends to saturate at high voltages. Higher charge events temporarily lower the electric field inside the holes, reducing the gain until the charge is evacuated. This field reduction causes a slower-than-exponential rise of the gain with voltage as is seen for the VERTEC glass and HPL Bakelite plates (figures \ref{fig:rpwell_gain_curves}c and \ref{fig:rpwell_gain_curves}d, respectively). This could also explain the gain saturation at earlier values at increased counting rates: although the gain is lower, charges accumulate faster (due to the rate), reducing the electric field within the hole. The RPWELL detectors with 2 mm and 4 mm thick Semitron plates (figures \ref{fig:gain_curves_semitron}b and \ref{fig:gain_curves_semitron}c respectively) only showed saturation at $\sim 10 \; \text{Hz/mm}^2$ at a gain of $3 \times 10^4$, while for the lower rates they did not saturate at the explored voltage values. In contrast, the response of the RPWELL with 0.6 mm Semitron plate (figure \ref{fig:gain_curves_semitron}a) did not saturate even at rates of $10^3 \; \text{Hz/mm}^2$ at a gain of $\sim 10^5$; although at lower voltage. Raising the voltage to that explored for the 2 mm and 4 mm Semitron plates was not possible due to the onset of current surges in the power supply's current monitor (discussed below)---although with no voltage drops.

For the THWELL and RWELL detectors the gain measurements were terminated at the appearance of occasional discharges (characterized by current spikes in the power supply, and voltage drops in the power supply). The gain measurements in all RPWELL detectors were terminated due to current surges in the power supply's current monitor but without any voltage drops. At $\sim 1050$ V (gain $\sim 10^5 $) across the RPWELL, the Semitron-plate detector developed a leakage current of $\sim 10$ nA, increasing with voltage to $\sim 50$ nA at 1075 V (gain $\sim 1.3 \times 10^5 $ ). In addition, occasional current surges of up to 400 nA were observed with no apparent drop in voltage. Both the leakage current and current spikes vanished below 950 V. At 1075 V the pulse-height spectrum was still above noise, however the resolution degraded to $\sim 50\% $ FWHM, as opposed to $ \sim 20\% $ at 850 V (gain $\sim 2 \times 10^3 $). In addition, when the voltage was increased from 1050 V to 1075 V, an abrupt drop in counting rate occurred: from a consistent $\sim 10^3 \; \text{Hz/mm}^2$ down to $\sim 30 \; \text{Hz/mm}^2$. This drop might have been due to a possible transition from a proportional to a streamer mode, observed in RPCs \cite{duerdoth_transition_1994, cardarelli_avalanche_1996}. While lowering the voltage restored the rate and eliminated the aforementioned current, the gain showed a $2-10$ fold drop---recovering only after several hours. Re-assembling the detector after several days restored the original gain at the same voltage.

\subsection{Rate dependence}
\label{subsec:results_rate_dependence}
\begin{figure}
\centering
\includegraphics[width=0.8\textwidth]{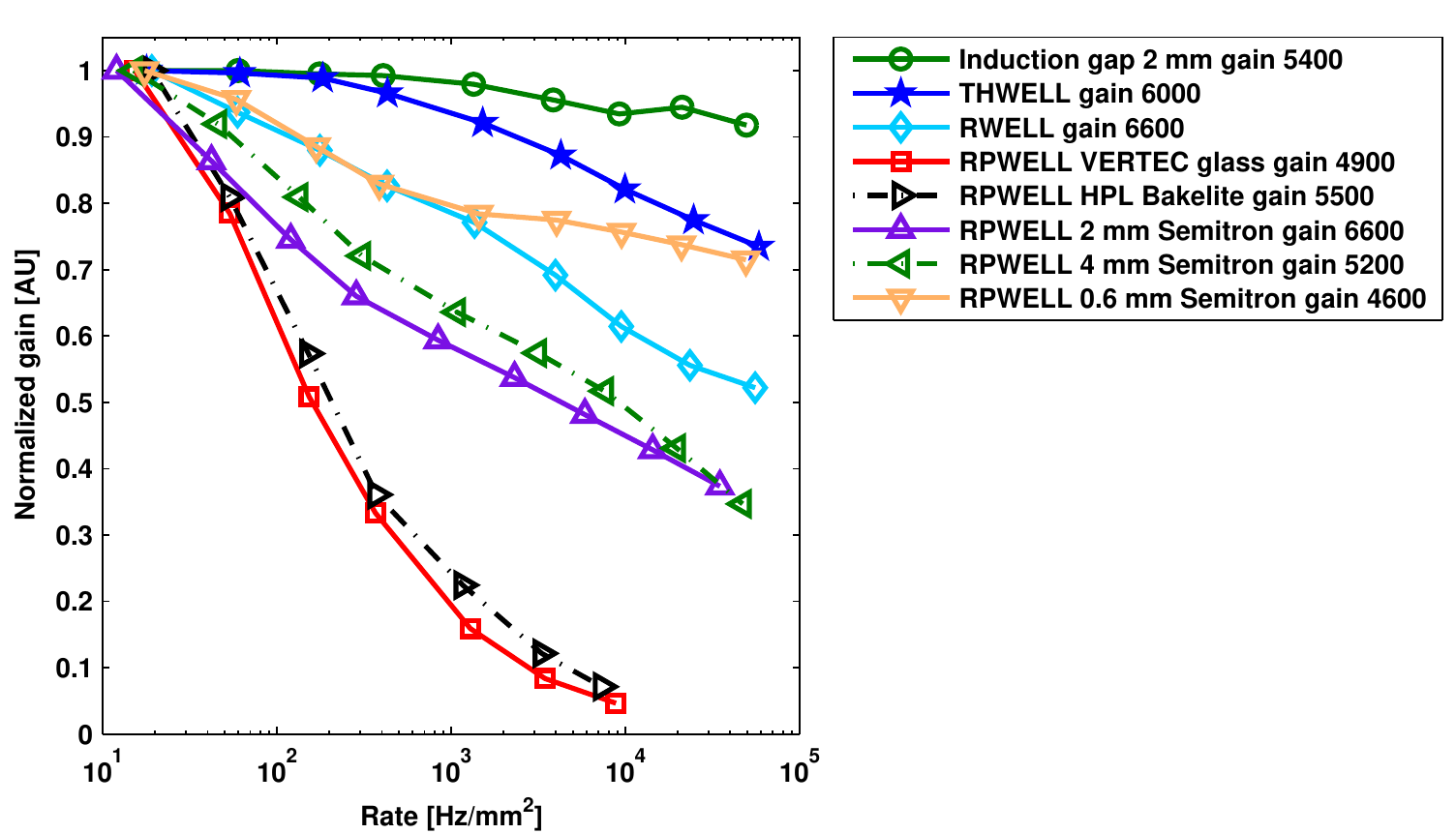}
\caption{Relative detector gain vs. x-rays rate. The different detector configurations were irradiated with a 1 mm diameter 8 keV x-ray beam. The initial gains were similar and are indicated in the figure.}
\label{fig:gain_vs_rate}
\end{figure}

The gain dependence on the counting rate (figure \ref{fig:gain_vs_rate}) was measured with the different detectors (configuration of figure \ref{fig:rpwell_configuration}); the gain variation was found to alter between the different anode materials. The RPWELL with VERTEC-glass and HPL-Bakelite had the worst performance, losing over 90\% of their original gain values over a 2 orders of magnitude increase in rate ($2\times10^1-4\times10^3$ Hz/mm$^2$). The 0.6 mm Semitron RPWELL ($2\times 10^9 \; \Omega$cm) performed slightly better than the RWELL detector (with 1 M$\Omega$/sq), losing $\sim 30\%$ of its gain; while the 2 mm and 4 mm Semitron RPWELL performed slightly worse, losing $\sim 60 \%$ of their gain over the same rate range ($2\times10^1-4\times10^3$ Hz/mm$^2$). The 0.4 mm thick THGEM with 2 mm induction gap was found to be the most robust to rate changes ($<10\%$ gain drop over the same range), in agreement with previous results \cite{peskov_further_2010}. The THWELL showed a $\sim 25\%$ drop in gain, however this might have been due also to its double thickness (0.8 mm) compared to that of the THGEM with the induction gap.

\subsection{Exposure to high primary charge}

\begin{figure}
\centering
\includegraphics[width=0.78\textwidth]{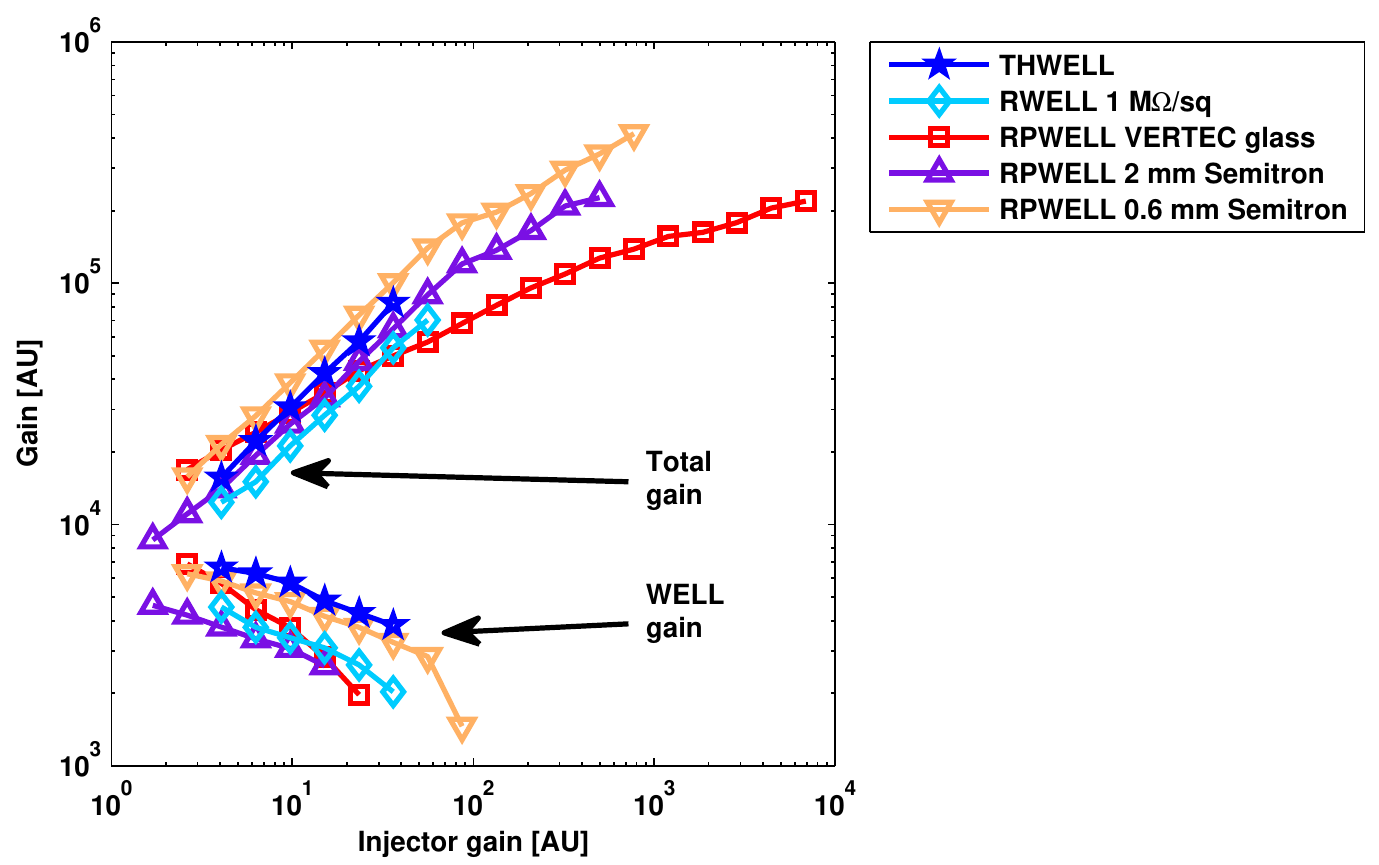}
\caption{Total gain of the double structure (figure \protect\ref{fig:injector_configuration}) and gain of the WELL element (high and low charge peaks in figure \protect\ref{fig:injector_spectrum} respectively). The initial gain of each WELL detector was set to $\sim 5\times10^3$. The rate was $\sim 10^{-1} \; \text{Hz/mm}^2$, with $\text{E}_\text{drift} = \text{E}_\text{trans} = 0.5 \; \text{kV/cm}$. }
\label{fig:gain_curves_injector}
\end{figure}

\begin{figure}
\centering
\begin{subfigure}{0.45\textwidth}
\includegraphics[width=\textwidth]{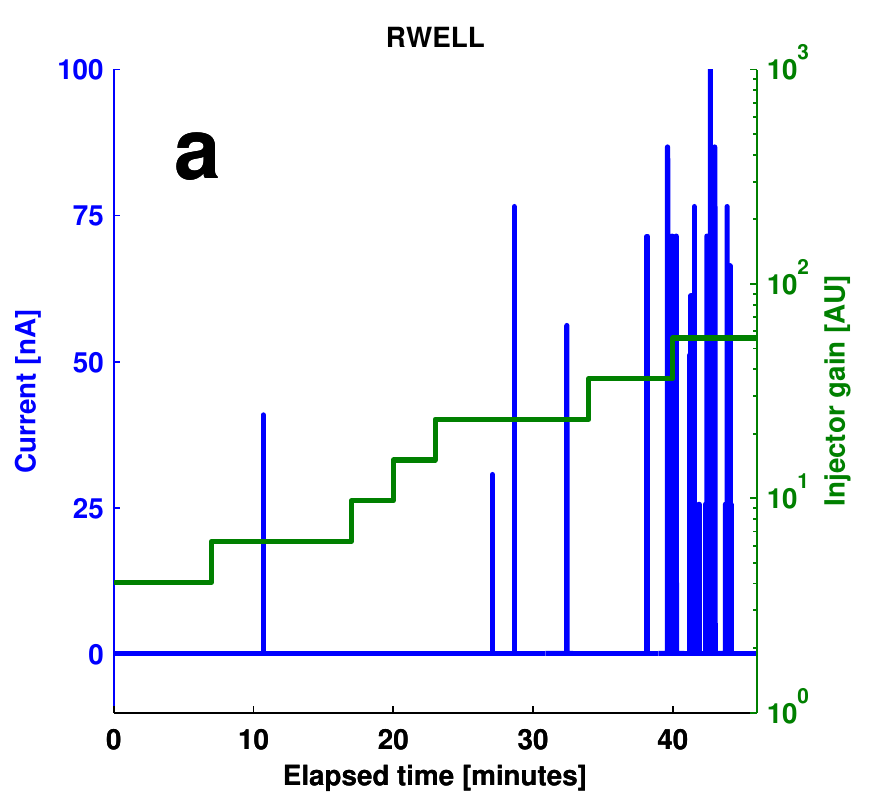}
\end{subfigure}
\begin{subfigure}{0.45\textwidth}
\includegraphics[width=\textwidth]{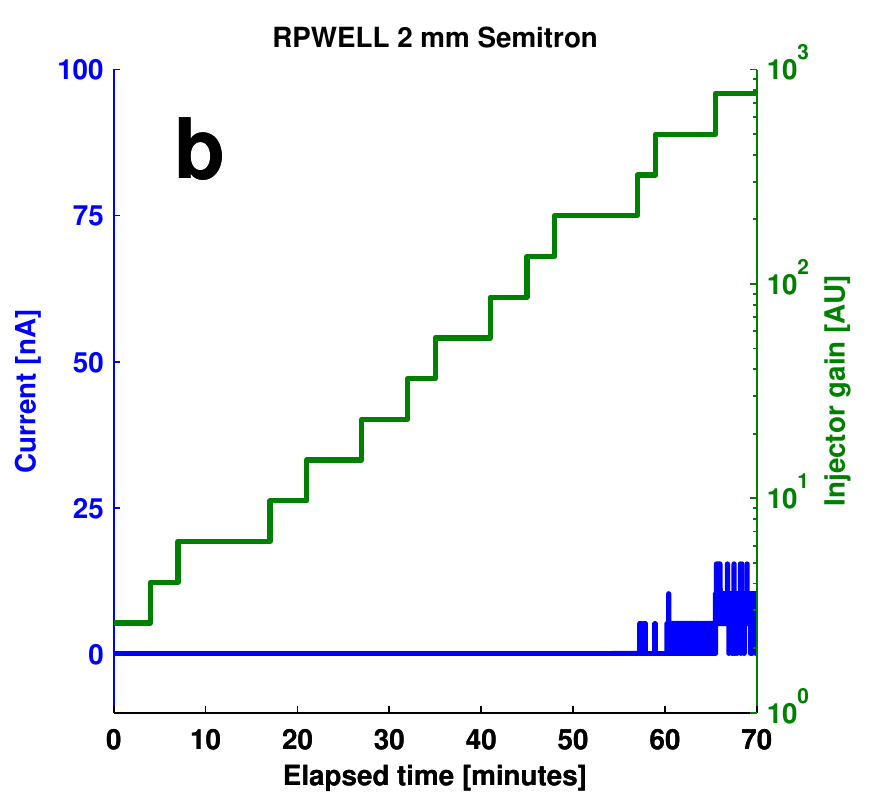}
\end{subfigure}
\caption{Current supplied to the top electrode and the injector gain for the RWELL (a) and RPWELL with 2 mm thick Semitron plate (b), for the setup shown in figure \protect\ref{fig:injector_configuration}. The current was measured from the current monitor of the  CAEN N471A power supply; the injector voltage was recorded manually.}
\label{fig:injector_imon}
\end{figure}

Figure \ref{fig:gain_curves_injector} shows the gain-response of the detector to highly ionizing events generated with the charge injector (figure \ref{fig:injector_configuration}). Each of the investigated WELL elements (THWELL, RWELL, RPWELL) was kept at a fixed voltage (resulting in a gain of $\sim 5000$), while the THGEM-injector gain was increased. Figure \ref{fig:injector_imon} shows the current supplied to the top of the WELL electrode and the injector gain for the assays involving the RWELL and RPWELL with the 2 mm thick Semitron plate. The THWELL and RWELL (figure \ref{fig:injector_imon}a) detectors experienced sparks---inducing voltage drops---at injector gains of $\sim 36$ and $\sim 56$ respectively. The RPWELLs (e.g. with 2 mm thick Semitron plate, figure \ref{fig:injector_imon}b) showed no sparks; however at injector gains above $\sim 10^2$, leakage currents were observed ($\sim 10 - 50$ nA) which depended on the injector gain; they vanished when the gain was reduced. The gain of all of the WELL-type detectors dropped with increasing the injector gain; this drop was not permanent; reducing the injector gain restored the original value. The total-gain vs. the injector-gain slopes are rather similar for  the THWELL, RWELL and RPWELL with the Semitron anode;  the slope with the RPWELL with VERTEC glass anode (highest resistivity, $\sim 8 \times 10^{12} \; \Omega$cm) showed a steeper decline. This may be due to an additional effect resulting from slower clearance of charges from the WELL holes, affecting the multiplication of the following event.

\section{Discussion} \label{sec:discussion}

We have investigated the properties of a novel Resistive-Plate WELL multiplier, RPWELL; it is a novel configuration in the THGEM family, aiming at a discharge-free operation under highly ionizing background, at moderate counting rates. The RPWELL investigated here, with soft x-rays in 1 atm Ne/5\%CH$_4$, comprised of a single-faced THGEM electrode, coupled to a metal anode via a resistive plate of high bulk resistivity. All RPWELL prototypes investigated here, with various resistive materials, yielded spectra above noise (figure \ref{fig:rpwell_spectra}); they provided discharge-free operation, with some dark-current ``activity'' at the higher voltage range but without any noticeable potential drops. Some of the resistive plates, like glass (of the highest bulk resistivity, $\sim 8 \times 10^{12} \; \Omega$cm), led to gain saturation at relatively low rates ($\sim 10$ Hz/mm$^2$); it is attributed to the slower charge evacuation causing field reduction within the holes. Among the different resistive materials investigated, 0.6 mm thick Semitron polymer, of a bulk resistivity $\sim 2 \times 10^9 \; \Omega$cm yielded the best results: a pulse-height resolution of $\sim 23\%$ FWHM (figure \ref{fig:rpwell_spectra}c) for 8 keV x-rays and gain stability (no discharges) up to values of $\sim 5 \times 10^4$ under counting rates of $\sim 10^3$ Hz/mm$^2$ (figure \ref{fig:gain_curves_semitron}a)---without charge saturation. Some hysteresis effects (pulse-height variations) were observed during the measurements, with the most resistive materials; these were not due to permanent damage to the detector and though being of key importance---their study is beyond the scope of this work, and requires more profound investigations.

The results of the rate-dependence of the pulse-height show that the RPWELL with the 0.6 mm thick Semitron polymer electrode has attractive qualities. Its rate-induced gain drop was slightly better than the RWELL  with $\sim$1 M$\Omega$/sq resistive film, with a $\sim 30\%$ pulse-height drop over a 3 orders of magnitude increase in rate: from 10 to 10$^4$ Hz/mm$^2$ (figure \ref{fig:gain_vs_rate});  conversely, it provided a fully spark-free operation and a high dynamic range. This result is similar to what was recently observed with a resistive MICROMEGAS detector \cite{wotschack_development_2012}.

The results obtained with the ``charge-injector'' (primary charge multiplier added to mimic highly ionizing events; figure \ref{fig:injector_configuration}), showed that the RPWELL configurations with the highly-resistive electrodes are robust relative to the THWELL (with metal anode) and RWELL (with $\sim$1 M$\Omega$/sq resistive film). The RPWELL configurations did not spark even at hundred-fold higher injected charges (reached with the injector; e.g. figure \ref{fig:injector_imon}). This indicates upon their potential robust operation in the presence of highly ionizing events (figure \ref{fig:gain_curves_injector}). The drop in gain due to the high injected charge, was larger for the electrodes of highest resistivity: glass (figure \ref{fig:gain_curves_injector}) and Bakelite (not shown in the figure).

Both the rate dependence of the gain and the drop in gain with highly ionizing events can be explained by the long ``time-constant'' of the highly resistive layers. It can be estimated by approximating our detector as a parallel plate, using

\begin{equation}
\tau \approx \rho \epsilon
\end{equation}

\noindent where $\tau$ is the time constant, $\rho$ is the bulk resistivity, and $\epsilon$ is the dielectric constant of the plate. Taking $\epsilon=4\epsilon_0$, where $\epsilon_0$ is the vacuum permittivity, gives $\tau \approx 4 \times 10^{-4} - 4\times 10^{-1}$ s for bulk resistivity $10^9 - 10^{12} \; \Omega$cm. Charges collected at the bottom of the hole on the resistive layer will reduce the field inside the hole, and subsequently reduce the gain. High rates and high charge will cause similar drops in gain, because they affect the rate of accumulation of charge on the layer. Lippmann et al. \cite{lippmann_rate_2006} have calculated field fluctuations on the order of $\sim 8\%$ due to incoming particles at 6 $\text{Hz/mm}^2$ for the RPC parallel plate geometry using quasi-static approximation of Maxwell’s equations.  Based on the gain curves measured (figures \ref{fig:rpwell_gain_curves} and \ref{fig:gain_curves_semitron}), at a gain of $5 \times 10^3$, a $0.7\%$ drop in the voltage, would cause a $\sim10\%$ drop in gain.  The electric field values within the THGEM holes are of the order of $\sim 10 \; \text{kV/cm}$. A charge as small as $\sim 35$ fC distributed uniformly at the bottom of the hole ($\sim 0.2 \; \text{mm}^2$), is sufficient to cause a $\sim 1\%$ drop in the field. This is a typical charge in the system: equivalent of an 8 keV x-ray conversion at a gain of $\sim 10^3$. This indicates that charge collecting on the resistive plate's surface can substantially distort the field inside the hole, and lower the effective gain.

Our preliminary results indicate that the RPWELL might become a very promising detector element. The thin 0.6 mm Semitron resistive plate provided so far the best performance in this operation mode; thinner plates have not yet been investigated. The current spikes that were observed with the resistive electrodes at very high gains were limited and did not trip the power supply. It is yet unclear whether these pulses were due to a transition from a proportional to Geiger mode of operation (similar to the streamer-mode operation of RPCs), or to a pre-onset of electrical breakdown in the material. However, it is important to stress that these effects appeared much above normal working conditions. In addition, the lack of a voltage drop in the power supply does not necessarily indicate that the detector has maintained its potential. We have shown that compared to the THWELL and RWELL, the power supply is stable, however efficiency studies are needed to assess the performance of the RPWELL in the face of current surges.

The materials investigated here (table \ref{tab:materials}) hardly exhaust the present list of commercially available anti-static and dissipative materials (both of the ion- and electron-conductive kinds). The electron-conductive materials have a more stable (in time, with current applied) bulk resistivity,\footnote{Private communication with Dr. P. Fonte.} making them more attractive for detector applications. Ceramics \cite{laso_ceramic_2012, naumann_ceramics_2011, garcia_extreme_2012} and doped glass \cite{wang_development_2010} have also drawn excitement within the RPC community, offering potential higher-rate operation. New materials, tailored for this application, can be engineered in collaboration with material scientists and industry.

While some of the principles of the RPWELL are similar to the RPC, the RPWELL offers some advantages over the geometry of the RPC. First, the region of multiplication is confined to the holes, the RPWELL can be preceded by large conversion/drift volumes while maintaining deposited energy proportionality. Second, the RPWELL can operate with standard counting gases, compared to RPCs, which require somewhat problematic gas mixtures \cite{peskov_challenges_2012, gonzalez-diaz_challenges_2013}). Moreover the RPWELL requires relatively low operation voltages. In addition, the RPWELL has a closed geometry, limiting avalanche divergence by photon feedback. Aging effects, which are of critical importance to RPCs, were not investigated in this work.

In conclusion, the RPWELL may prove to be an important step forward in the evolution of MPGDs. It has been shown that some of the present RPWELL configurations have a broad dynamic range, and operate in a stable way, at reasonable counting rates, at the presence of highly ionizing events. This thin single-element detector may have numerous applications; an important one would be in future semi-Digital \cite{linssen_physics_2012}  or Digital Hadronic Calorimetry (DHCAL \cite{brau_international_2013}), where thin, robust detectors of large dynamic range are necessary. We have shown that despite the introduction of highly resistive materials, the rate capabilities of the RPWELL are adequate for this application in future-collider experiments. Studies for assessing its detection efficiency and multiplicity, similar to that performed with THGEM and SRWELL detectors \cite{bressler_beam_2013,arazi_beam_2013} are in preparation.

\section*{Acknowledgments}
We would like to thank Drs.  F. Sauli, P. Fonte and G. Sekhniaidze for supplying the resistive-material samples investigated in this work. We would also like to thank Dr. V. Dangendorf for fruitful discussions, and Dr. S. Shilstein for technical assistance. This work was supported in part by the Israel-USA Binational Science Foundation (Grant 2008246). A. Breskin is the W.P. Reuther Professor of Research in the Peaceful use of Atomic Energy.

\bibliography{RPWELL}

\providecommand{\href}[2]{#2}\begingroup\raggedright\begin{thebibliography}{10}

\bibitem{nanjo_neutron_2005}
H.~Nanjo, T.~Bando, K.~Hasuko, M.~Ishino, T.~Kobayashi, T.~Takemoto et~al.,
  \emph{Neutron sensitivity of thin gap chambers},
  \href{http://dx.doi.org/10.1016/j.nima.2004.11.052}{\emph{Nucl. Instrum.
  Meth. A} {\bf 543} (May, 2005) 441--453}.

\bibitem{adam_effect_2005}
W.~Adam, T.~Bergauer, M.~Friedl, R.~Fruehwirth, J.~Hrubec, M.~Krammer et~al.,
  \emph{The effect of highly ionising particles on the {CMS} silicon strip
  tracker}, \href{http://dx.doi.org/10.1016/j.nima.2004.11.049}{\emph{Nucl.
  Instrum. Meth. A} {\bf 543} (May, 2005) 463--482}.

\bibitem{sauli_gem:_1997}
F.~Sauli, \emph{{GEM:} a new concept for electron amplification in gas
  detectors},
  \href{http://dx.doi.org/10.1016/S0168-9002(96)01172-2}{\emph{Nucl. Instrum.
  Meth. A} {\bf 386} (Feb., 1997) 531--534}.

\bibitem{sauli_imaging_2007}
F.~Sauli, \emph{Imaging with the gas electron multiplier},
  \href{http://dx.doi.org/10.1016/j.nima.2007.06.100}{\emph{Nucl. Instrum.
  Meth. A} {\bf 580} (Oct., 2007) 971--973}.

\bibitem{chechik_thick_2005}
R.~Chechik, A.~Breskin and C.~Shalem, \emph{Thick {GEM-like} multipliers—a
  simple solution for large area {UV-RICH} detectors},
  \href{http://dx.doi.org/10.1016/j.nima.2005.08.003}{\emph{Nucl. Instrum.
  Meth. A} {\bf 553} (Nov., 2005) 35--40}.

\bibitem{duval_hybrid_2012}
S.~Duval, L.~Arazi, A.~Breskin, R.~Budnik, W.-T. Chen, H.~Carduner et~al.,
  \emph{Hybrid multi micropattern gaseous photomultiplier for detection of
  liquid-xenon scintillation},
  \href{http://dx.doi.org/10.1016/j.nima.2011.11.018}{\emph{Nucl. Instrum.
  Meth. A} {\bf 695} (Dec., 2012) 163--167}.

\bibitem{alexopoulos_spark-resistant_2011}
T.~Alexopoulos, J.~Burnens, R.~de~Oliveira, G.~Glonti, O.~Pizzirusso,
  V.~Polychronakos et~al., \emph{A spark-resistant bulk-micromegas chamber for
  high-rate applications},
  \href{http://dx.doi.org/10.1016/j.nima.2011.03.025}{\emph{Nucl. Instrum.
  Meth. A} {\bf 640} (June, 2011) 110--118}.

\bibitem{wotschack_development_2012}
J.~Wotschack, \emph{Development of micromegas muon chambers for the {ATLAS}
  upgrade},
  \href{http://dx.doi.org/10.1088/1748-0221/7/02/C02021}{\emph{{JINST}} {\bf 7}
  (Feb., 2012) C02021--C02021}.

\bibitem{oliveira_first_2007}
R.~Oliveira, V.~Peskov, F.~Pietropaolo and P.~Picchi, \emph{First tests of
  thick {GEMs} with electrodes made of a resistive kapton},
  \href{http://dx.doi.org/10.1016/j.nima.2007.03.010}{\emph{Nucl. Instrum.
  Meth. A} {\bf 576} (June, 2007) 362--366}.

\bibitem{fonte_development_2012}
P.~Fonte, E.~Nappi, P.~Martinengo, R.~Oliveira, V.~P.~F. Pietropaolo and
  P.~Picchi, \emph{Development and preliminary tests of resistive microdot and
  microstrip detectors}, {\emph{{arXiv:1203.3658}} (Mar., 2012) }.

\bibitem{peskov_advances_2009}
V.~Peskov, \emph{Advances in micro-pattern gaseous detectors and their
  applications}, {\emph{{arXiv:0906.5215}} (June, 2009) }.

\bibitem{peskov_advances_2012}
V.~Peskov, P.~Fonte, P.~Martinengo, E.~Nappi, R.~Oliveira, F.~Pietropaolo
  et~al., \emph{Advances in the development of micropattern gaseous detectors
  with resistive electrodes},
  \href{http://dx.doi.org/10.1016/j.nima.2010.09.171}{\emph{Nucl. Instrum.
  Meth. A} {\bf 661, Supplement 1} (Jan., 2012) S153--S155}.

\bibitem{shalem_advances_2006-1}
C.~Shalem, R.~Chechik, A.~Breskin and K.~Michaeli, \emph{Advances in thick
  {GEM-like} gaseous electron {multipliers—Part} i: atmospheric pressure
  operation}, \href{http://dx.doi.org/10.1016/j.nima.2005.12.241}{\emph{Nucl.
  Instrum. Meth. A} {\bf 558} (Mar., 2006) 475--489}.

\bibitem{cortesi_thgem_2009}
M.~Cortesi, V.~Peskov, G.~Bartesaghi, J.~Miyamoto, S.~Cohen, R.~Chechik et~al.,
  \emph{{THGEM} operation in ne and {Ne/CH4}},
  \href{http://dx.doi.org/10.1088/1748-0221/4/08/P08001}{\emph{{JINST}} {\bf 4}
  (Aug., 2009) P08001--P08001}.

\bibitem{breskin_concise_2009}
A.~Breskin, R.~Alon, M.~Cortesi, R.~Chechik, J.~Miyamoto, V.~Dangendorf et~al.,
  \emph{A concise review on {THGEM} detectors},
  \href{http://dx.doi.org/10.1016/j.nima.2008.08.062}{\emph{Nucl. Instrum.
  Meth. A} {\bf 598} (Jan., 2009) 107--111}.

\bibitem{di_mauro_development_2007}
A.~Di~Mauro, B.~Lund-Jensen, P.~Martinengo, E.~Nappi, R.~Oliveira, V.~Peskov
  et~al., \emph{Development of innovative micro-pattern gaseous detectors with
  resistive electrodes and first results of their applications},
  \href{http://dx.doi.org/10.1016/j.nima.2007.07.083}{\emph{Nucl. Instrum.
  Meth. A} {\bf 581} (Oct., 2007) 225--231}.

\bibitem{bartol_c..t._1996}
F.~Bartol, M.~Bordessoule, G.~Chaplier, M.~Lemonnier and S.~Megtert, \emph{The
  {C.A.T.} pixel proportional gas counter detector},
  \href{http://dx.doi.org/10.1051/jp3:1996127}{\emph{J. Phys. {III}} {\bf 6}
  (Mar., 1996) 337--347}.

\bibitem{bellazzini_well_1999}
R.~Bellazzini, M.~Bozzo, A.~Brez, G.~Gariano, L.~Latronico, N.~Lumb et~al.,
  \emph{The {WELL} detector},
  \href{http://dx.doi.org/10.1016/S0168-9002(98)01187-5}{\emph{Nucl. Instrum.
  Meth. A} {\bf 423} (Feb., 1999) 125--134}.

\bibitem{rocco_development_2010}
E.~Rocco, \emph{Development of a gaseous photon detector for Cherenkov imaging
  applications}.
\newblock {PhD} thesis, Universit`a degli Studi di Torino, Torino, Italy, 2010.

\bibitem{alfonsi_performance_2009}
M.~Alfonsi, G.~Croci, S.~D. Pinto, R.~d. Oliveira, P.~Picchi, E.~Rocco et~al.,
  \emph{Performance measurements on closed-geometry, {GEM-like} detectors},  in
  \emph{1st international conference on Micro Pattern Gaseous Detectors
  {MPGD2009}, unpublished}, (Kolympari Crete Greece), June, 2009.

\bibitem{amram_position_2011}
N.~Amram, G.~Bella, Y.~Benhammou, M.~A. Diaz, E.~Duchovni, E.~Etzion et~al.,
  \emph{Position resolution and efficiency measurements with large scale thin
  gap chambers for the super {LHC}},
  \href{http://dx.doi.org/10.1016/j.nima.2010.06.311}{\emph{Nucl. Instrum.
  Meth. A} {\bf 628} (Feb., 2011) 177--181}.

\bibitem{arazi_beam_2013}
L.~Arazi, C.~D.~R. Azevedo, A.~Breskin, S.~Bressler, L.~Moleri, H.~N. da~Luz
  et~al., \emph{Beam studies of the segmented resistive {WELL:} a potential
  thin sampling element for digital hadron calorimetry},
  {\emph{{arXiv:1305.1585}} (May, 2013) }.

\bibitem{bressler_beam_2013}
S.~Bressler, L.~Arazi, H.~N. da~Luz, C.~D.~A. Azevedo, L.~Moleri, E.~Oliveri
  et~al., \emph{Beam studies of novel {THGEM-based} potential sampling elements
  for digital hadron calorimetry}, {\emph{{arXiv:1305.4657}} (May, 2013) }.

\bibitem{arazi_laboratory_2013}
L.~Arazi, M.~Pitt, S.~Bressler, L.~Moleri, A.~Rubin, S.~Shilstein et~al.,
  \emph{Laboratory studies of {THGEM-based} {WELL} structures with resistive
  anode}, {\emph{in preparation} (2013) }.

\bibitem{arazi_thgem-based_2012}
L.~Arazi, H.~Natal~da Luz, D.~Freytag, M.~Pitt, C.~D.~R. Azevedo, A.~Rubin
  et~al., \emph{{THGEM-based} detectors for sampling elements in {DHCAL:}
  laboratory and beam evaluation},
  \href{http://dx.doi.org/10.1088/1748-0221/7/05/C05011}{\emph{{JINST}} {\bf 7}
  (May, 2012) C05011--C05011}.

\bibitem{santonico_development_1981}
R.~Santonico and R.~Cardarelli, \emph{Development of resistive plate counters},
  \href{http://dx.doi.org/10.1016/0029-554X(81)90363-3}{\emph{Nucl. Instrum.
  Meth. in Physics Research} {\bf 187} (Aug., 1981) 377--380}.

\bibitem{laso_ceramic_2012}
A.~Laso, \emph{Ceramic resistive plate chambers for high rate environments},
  {\emph{{PoS} ({RPC2012)}} {\bf 66} (2012) }.

\bibitem{naumann_ceramics_2011}
L.~Naumann, R.~Kotte, D.~Stach and J.~Wüstenfeld, \emph{Ceramics high rate
  timing {RPC}},
  \href{http://dx.doi.org/10.1016/j.nima.2010.06.302}{\emph{Nucl. Instrum.
  Meth. A} {\bf 628} (Feb., 2011) 138--141}.

\bibitem{garcia_extreme_2012}
A.~L. Garcia, M.~Kaspar, B.~Kämpfer, R.~Kotte, L.~Naumann, R.~Peschke et~al.,
  \emph{Extreme high-rate capable timing resistive plate chambers with ceramic
  electrodes},
  \href{http://dx.doi.org/10.1088/1748-0221/7/10/P10012}{\emph{{JINST}} {\bf 7}
  (Oct., 2012) P10012--P10012}.

\bibitem{wang_development_2010}
J.~Wang, Y.~Wang, X.~Zhu, W.~Ding, Y.~Li, J.~Cheng et~al., \emph{Development of
  multi-gap resistive plate chambers with low-resistive silicate glass
  electrodes for operation at high particle fluxes and large transported
  charges}, \href{http://dx.doi.org/10.1016/j.nima.2010.04.056}{\emph{Nucl.
  Instrum. Meth. A} {\bf 621} (Sept., 2010) 151--156}.

\bibitem{gonzalez-diaz_challenges_2013}
D.~Gonzalez-Diaz and A.~Sharma, \emph{Challenges for resistive gaseous
  detectors towards {RPC2014}},
  \href{http://dx.doi.org/10.1088/1748-0221/8/02/T02001}{\emph{{JINST}} {\bf 8}
  (Feb., 2013) T02001}.

\bibitem{bashkirov_novel_2009}
V.~A. Bashkirov, R.~F. Hurley and R.~W. Schulte, \emph{A novel detector for
  {2D} ion detection in low-pressure gas and its applications},  in \emph{2009
  {IEEE} Nuclear Science Symposium Conference Record ({NSS/MIC)}},
  pp.~694--698, {IEEE}, Oct., 2009.
\newblock \href{http://dx.doi.org/10.1109/NSSMIC.2009.5402061}{DOI}.

\bibitem{di_mauro_new_2006}
A.~Di~Mauro, B.~Lund-Jensen, P.~Martinengo, E.~Nappi, V.~Peskov, L.~Periale
  et~al., \emph{A new {GEM-like} imaging detector with electrodes coated with
  resistive layers},  in \emph{{IEEE} Nuclear Science Symposium Conference
  Record, 2006}, vol.~6, pp.~3852--3859, Nov., 2006.
\newblock \href{http://dx.doi.org/10.1109/NSSMIC.2006.353831}{DOI}.

\bibitem{meghna_measurement_2012}
K.~K. Meghna, A.~Banerjee, S.~Biswas, S.~Bhattacharya, S.~Bose,
  S.~Chattopadhyay et~al., \emph{Measurement of electrical properties of
  electrode materials for the bakelite resistive plate chambers},
  \href{http://dx.doi.org/10.1088/1748-0221/7/10/P10003}{\emph{{JINST}} {\bf 7}
  (Oct., 2012) P10003}.

\bibitem{national_instruments_labview_2012}
N.~Instruments, \emph{{LabVIEW} {SignalExpress}},  2012.

\bibitem{moleri_investigation_2013}
L.~Moleri, S.~Bressler and A.~Breskin, \emph{Investigation of {THGEM}
  structures over a broad dynamic range}, {\emph{in preparation} (2013) }.

\bibitem{duerdoth_transition_1994}
I.~Duerdoth, S.~Clowes, J.~Freestone, F.~Loebinger, J.~Lomas, S.~Snow et~al.,
  \emph{The transition from proportional to streamer mode in a resistive plate
  chamber}, \href{http://dx.doi.org/10.1016/0168-9002(94)90751-X}{\emph{Nucl.
  Instrum. Meth. A} {\bf 348} (Sept., 1994) 303--306}.

\bibitem{cardarelli_avalanche_1996}
R.~Cardarelli, V.~Makeev and R.~Santonico, \emph{Avalanche and streamer mode
  operation of resistive plate chambers},
  \href{http://dx.doi.org/10.1016/S0168-9002(96)00811-X}{\emph{Nucl. Instrum.
  Meth. A} {\bf 382} (Nov., 1996) 470--474}.

\bibitem{peskov_further_2010}
V.~Peskov, M.~Cortesi, R.~Chechik and A.~Breskin, \emph{Further evaluation of a
  {THGEM} {UV-photon} detector for {RICH} – comparison with {MWPC}},
  \href{http://dx.doi.org/10.1088/1748-0221/5/11/P11004}{\emph{{JINST}} {\bf 5}
  (Nov., 2010) P11004--P11004}.

\bibitem{lippmann_rate_2006}
C.~Lippmann, W.~Riegler and A.~Kalweit, \emph{Rate effects in resistive plate
  chambers},
  \href{http://dx.doi.org/10.1016/j.nuclphysbps.2006.07.037}{\emph{Nuclear
  Physics B - Proceedings Supplements} {\bf 158} (Aug., 2006) 127--130}.

\bibitem{peskov_challenges_2012}
V.~Peskov, \emph{Challenges for {RPCs} and resistive micropattern detectors in
  the next few years}, {\emph{{arXiv:1204.2144}} (Apr., 2012) }.

\bibitem{linssen_physics_2012}
L.~Linssen, A.~Miyamoto, M.~Stanitzki and H.~Weerts, \emph{Physics and
  detectors at {CLIC:} {CLIC} conceptual design report},
  {\emph{{arXiv:1202.5940}} (Feb., 2012) }.

\bibitem{brau_international_2013}
J.~Brau, Y.~Okada and N.~Walker, \emph{International linear collider reference
  design report, the international linear collider technical design report},
  tech. rep., {ILC} webplage, 2013.

\end{thebibliography}\endgroup
\bibliographystyle{MJ}

\end{document}